\documentclass[aps,prl,twocolumn,superscriptaddress]{revtex4-2}
\pdfoutput=1
\usepackage{graphicx}
\usepackage{amsmath}
\usepackage{amssymb}
\usepackage{siunitx}
\usepackage[version=4]{mhchem}
\usepackage{lmodern}
\usepackage{gensymb}
\usepackage{xcolor}
\usepackage{color}

\newcommand{\PP}{P\textsubscript{in}-P\textsubscript{out}}

\newcommand{\SiSo}{S\textsubscript{in}-S\textsubscript{out}}
\newcommand{\cuone}{Cu\textsubscript{I}}
\newcommand{\cutwo}{Cu\textsubscript{II}}
\newcommand{\tnone}{\textit{T}\textsubscript{N,I}}
\newcommand{\tntwo}{\textit{T}\textsubscript{N,II}}
\newcommand{\treor}{\textit{T}\textsubscript{R}}

\begin{document}

\righthyphenmin=4
\lefthyphenmin=4

\title{Direct visualization and control of antiferromagnetic domains and spin reorientation in a parent cuprate}

\author{K. L. Seyler}
\affiliation{
 Department of Physics, California Institute of Technology, Pasadena, CA 91125, USA
}
\affiliation{
 Institute for Quantum Information and Matter, California Institute of Technology, Pasadena, CA 91125, USA
}

\author{A. Ron}
\affiliation{
 Department of Physics, California Institute of Technology, Pasadena, CA 91125, USA
}
\affiliation{
 Institute for Quantum Information and Matter, California Institute of Technology, Pasadena, CA 91125, USA
}
\affiliation{
 Raymond and Beverly Sackler School of Physics and Astronomy, Tel-Aviv University, Tel Aviv, 69978, Israel
}

\author{D. Van Beveren}
\affiliation{
 Department of Physics, California Institute of Technology, Pasadena, CA 91125, USA
}
\affiliation{
 Institute for Quantum Information and Matter, California Institute of Technology, Pasadena, CA 91125, USA
} 
 
\author{C. R. Rotundu}
\affiliation{
 Stanford Institute for Materials and Energy Sciences, SLAC National Accelerator Laboratory, 2575 Sand Hill Road, Menlo Park, CA 94025, USA
}
 
 \author{Y. S. Lee}
\affiliation{
 Stanford Institute for Materials and Energy Sciences, SLAC National Accelerator Laboratory, 2575 Sand Hill Road, Menlo Park, CA 94025, USA
}
\affiliation{
 Department of Applied Physics, Stanford University, Stanford, CA 94305, USA
}
 
\author{D. Hsieh}
\affiliation{
 Department of Physics, California Institute of Technology, Pasadena, CA 91125, USA
}
\affiliation{
 Institute for Quantum Information and Matter, California Institute of Technology, Pasadena, CA 91125, USA
}

\begin{abstract}

We report magnetic optical second-harmonic generation (SHG) polarimetry and imaging on \ce{Sr2Cu3O4Cl2}, which allows direct visualization of the mesoscopic antiferromagnetic (AFM) structure of a parent cuprate. Temperature- and magnetic-field-dependent SHG reveals large domains with 90\degree{} relative orientations that are stabilized by a combination of uniaxial magnetic anisotropy and the Earth's magnetic field. Below a temperature \treor{} $\sim$ 97 K, we observe an unusual 90\degree{} spin reorientation transition, possibly driven by competing magnetic anisotropies of the two copper sublattices, which swaps the AFM domain states while preserving the domain structure. This allows deterministic switching of the AFM states by thermal or laser heating. Near \treor, the domain walls become exceptionally responsive to an applied magnetic field, with the Earth's field sufficient to completely expel them from the crystal. Our findings unlock opportunities to study the mesoscopic AFM behavior of parent cuprates and explore their potential for AFM technologies.

\end{abstract}

\maketitle
\date{\today}

Antiferromagnetic (AFM) materials host a rich variety of magnetic phenomena and are appealing for robust high-speed spin-based technologies \cite{Jungwirth2016-po, Baltz2018-he, Nemec2018-mm, Gomonay2018-mb}. Cuprate Mott insulators, the parent compounds of high-T$_c$ superconductors, are particularly intriguing AFM materials owing to their model Heisenberg behavior, record-high exchange interactions, and tunability with doping \cite{Manousakis1991-vx, Lee2006-co}. However, there is limited understanding of their mesoscopic magnetic properties due to the difficulty of achieving local readout of AFM order and spatial mapping of AFM domain wall distributions \cite{Parks2001-hr, Nafradi2016-zy}. Here we directly visualize AFM domains in the parent cuprate \ce{Sr2Cu3O4Cl2} using optical second-harmonic generation (SHG) polarimetry and imaging. We uncover a spin-reorientation transition that enables thermally controlled deterministic \ang{90} switching of AFM states and complete expulsion of AFM domain walls with Oersted-level magnetic fields.

Magnetic crystals that break time-reversal symmetry permit time-noninvariant ($c$-type) SHG processes that directly couple to the magnetic order parameter \cite{Fiebig2005-ms,Kirilyuk2005-lj}, making SHG a potentially powerful probe of AFM domains \cite{Fiebig2005-ms, Cheong2020-gi} and dynamics \cite{Fiebig2008-wn}. Although $c$-type SHG is most widely reported in the electric-dipole channel from noncentrosymmetric AFM materials \cite{Chauleau2017-td,Tzschaschel2019-eo,Sun2019-na,Chu2020-dm}, it has also been detected in weaker magnetic-dipole (MD) channels from centrosymmetric materials \cite{Kaminski2009-eh, Matsubara2010-li}. However, ideal AFM-ordered parent cuprates preserve time-reversal symmetry because even though time-reversal is locally broken at each Cu site, it is restored upon translation by a primitive lattice vector. This leads to perfect cancellation of $c$-type SHG radiation from the two magnetic sublattices. Therefore, cuprate antiferromagnetism is expected to be SHG inactive, as was recently confirmed in the prototypical compound \ce{Sr2CuO2Cl2} \cite{De_la_Torre2021-cj, De_la_Torre2021-ks}.

\begin{figure*}[!htb]
\includegraphics{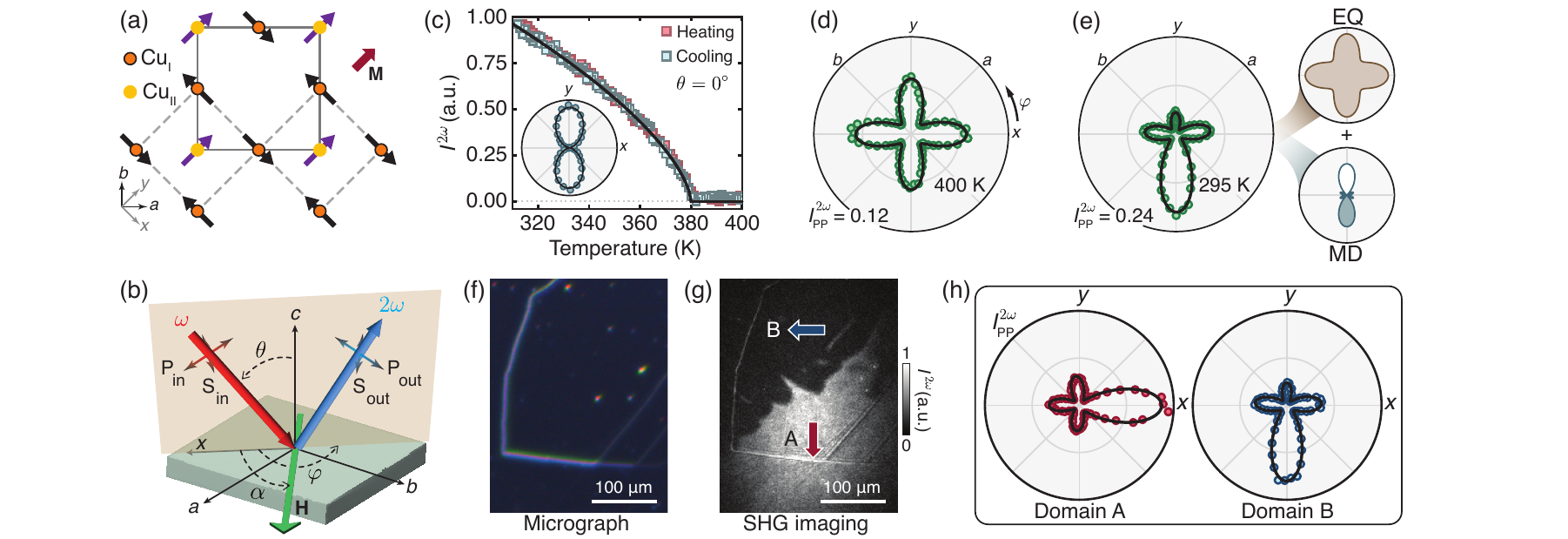}
\caption{\label{fig:m1} Local AFM readout in \ce{Sr2Cu3O4Cl2}. (a) Crystal and magnetic structure of single Cu-O layer in \ce{Sr2Cu3O4Cl2}. Only the moment induced by internal pseudodipolar field is depicted. Thick arrow indicates $\mathbf{M}$. (b) Schematic of RA-SHG experimental geometry, where angle of incidence ($\theta$), scattering plane angle ($\varphi$), in-plane magnetic field ($\mathbf{H}$) direction ($\alpha$), and input and output electric field polarizations (P or S) are varied. (c) Temperature dependence of normal incidence ($\theta = \ang{0}$) SHG intensity. Solid line is a least-squares fit to $I^{2\omega}\propto(\tnone-T)^{2\beta}$, where $\beta=0.32(3)$ and $\tnone=\SI{380(1)}{\kelvin}$. Fit is performed near \tnone{} ($\SI{360}{\kelvin}\leq T \leq \SI{380}{\kelvin}$) with uncertainties given as 1 standard deviation. Inset: normal incidence RA-SHG for co-linearly polarized excitation and detection beams measured at $T=\SI{295}{\kelvin}$ and fit by $\chi^{\text{MD}(\mathit{c})}$ ($mm'm'$) process (solid line). (d) Oblique-incidence ($\theta=\ang{10}$) \PP{} RA-SHG pattern at \SI{400}{\kelvin} fit by a $\chi^{\text{EQ}(\mathit{i})}$ ($4/mmm$) process (solid line). (e) \PP{} RA-SHG pattern ($\theta=\ang{10}$) at \SI{295}{\kelvin} fit to a coherent superposition of $\chi^{\text{EQ}(\mathit{i})}$ ($4/mmm$) and $\chi^{\text{MD}(\mathit{c})}$ ($mm'm'$) processes. EQ and MD processes are illustrated on the right, where patterns represent the P\textsubscript{in}-light-induced nonlinear polarization projected along P\textsubscript{out}. Filled and white lobes indicate opposite phase. (f) Dark-field optical micrograph of cleaved (001) \ce{Sr2Cu3O4Cl2}. Bright and dark lines correspond to surface terrace steps. (g) Wide-field SHG image under horizontal excitation polarization (along $x$ axis) at \SI{295}{\kelvin}. Domains A and B are labeled with arrow corresponding to $\mathbf{M}$. (h) \PP{} RA-SHG patterns at \SI{295}{\kelvin} for domains A and B.}
\end{figure*}

The centrosymmetric tetragonal structure (point group, $4/mmm$) of \ce{Sr2Cu3O4Cl2} is nearly identical to \ce{Sr2CuO2Cl2} except for an additional set of \ce{Cu^2+} ions (\cutwo) located in every other plaquette of the conventional \ce{CuO2} lattice (\cuone, Fig.~\ref{fig:m1}a) \cite{Grande1976-gm}. The \cuone{} spins interact via strong intralayer AFM exchange ($J\textsubscript{I} = \SI{130}{\meV}$) and order below ${\tnone\approx\SI{380}{\kelvin}}$, well above the AFM ordering temperature of the \cutwo{} sublattice (${\tntwo\approx\SI{40}{\kelvin}}$) \cite{Yamada1995-ar}. However, because \cutwo{} breaks the equivalence of neighboring \cuone{} sites, the AFM ordered \cuone{} sublattice becomes SHG active below \tnone. The AFM ordered \cuone{} sublattice generates a net field at the \cutwo{} sites via a weak pseudodipolar interaction \cite{Chou1997-fs, Kastner1999-ka}. This induces a polarization of \cutwo{} spins and slight canting of \cuone{} spins, resulting in a centrosymmetric AFM structure (point group $mm'm'$) with a small net in-plane ferromagnetic moment $\mathbf{M}$ \cite{Chou1997-fs, Kastner1999-ka} (Fig.~\ref{fig:m1}a). Four degenerate 90$^{\circ}$-rotated AFM domain configurations correspond to $\mathbf{M}$ along $[110]$, $[\bar{1}10]$, $[\bar{1}\bar{1}0]$ or $[1\bar{1}0]$, which can in principle be distinguished via MD SHG. A previous study used bulk magnetometry to infer the existence of stable AFM domains with \ang{90} relative orientations and a domain wall phase transition near \SI{100}{\kelvin} \cite{Parks2001-hr}, where it was proposed that domains are stabilized by entropic \cite{Parks2001-hr} or magnetoelastic \cite{Gomonay2011-im} effects. However, direct observation of AFM domains has remained elusive.

To establish the existence of an SHG response that directly couples to the magnetic order parameter, which can be represented by $\mathbf{M}$, we performed rotational anisotropy (RA) measurements on (001)-cleaved single crystals of \ce{Sr2Cu3O4Cl2} using a fast-rotating scattering-plane-based technique (Fig.~\ref{fig:m1}b) \cite{Harter2015-iy}. Under normal incidence ($\theta=\ang{0}$), a nonzero SHG signal appears below \tnone{} and shows no thermal hysteresis (Fig.~\ref{fig:m1}c), consistent with a continuous AFM transition. The dumbbell-shaped RA patterns are well described by a magnetization-induced MD process $P^{2\omega}_{i}=\chi_{ijk}^\text{MD(\textit{c})}E^{\omega}_{j}H^{\omega}_{k}$, where $\chi_{ijk}^\text{MD(\textit{c})}$ is an axial $c$-type susceptibility tensor respecting $mm'm'$ symmetry \cite{SI} \nocite{Noro1994-ba,Birss1964-bk,Reif1991-dx,Gomonay2007-ty} that relates the incident electric and magnetic fields at frequency $\omega$ to the induced polarization at 2$\omega$, and the subscripts run through $x$, $y$ and $z$. Below \tnone{}, $\chi^\text{MD(\textit{c})}$ exhibits a power-law temperature dependence with a fitted critical exponent $\beta = 0.32(3)$, which is consistent with the critical exponent of the staggered and saturated moments measured by neutron diffraction \cite{Yamada1995-ar,Kim2001-fz} and magnetometry \cite{Chou1997-fs,Kastner1999-ka} to within experimental errors. Together, these data confirm a $\chi_{ijk}^\text{MD(\textit{c})}$ response that scales linearly with $\mathbf{M}$. To determine the sign of $\chi_{ijk}^\text{MD(\textit{c})}$, we measured RA patterns at oblique incidence ($\theta=\ang{10}$), where a temperature-independent time-invariant ($i$-type) electric quadrupole (EQ) SHG process becomes active (Fig.~\ref{fig:m1}d). Below \tnone, the EQ and MD terms interfere to produce an RA pattern with broken rotational symmetry, revealing the sign of $\chi_{ijk}^\text{MD(\textit{c})}$ (Fig.~\ref{fig:m1}e, \cite{SI}). All four AFM domain configurations can therefore be locally read out from the orientation of the large lobe in the RA pattern.

\begin{figure*}[!htb]
\includegraphics{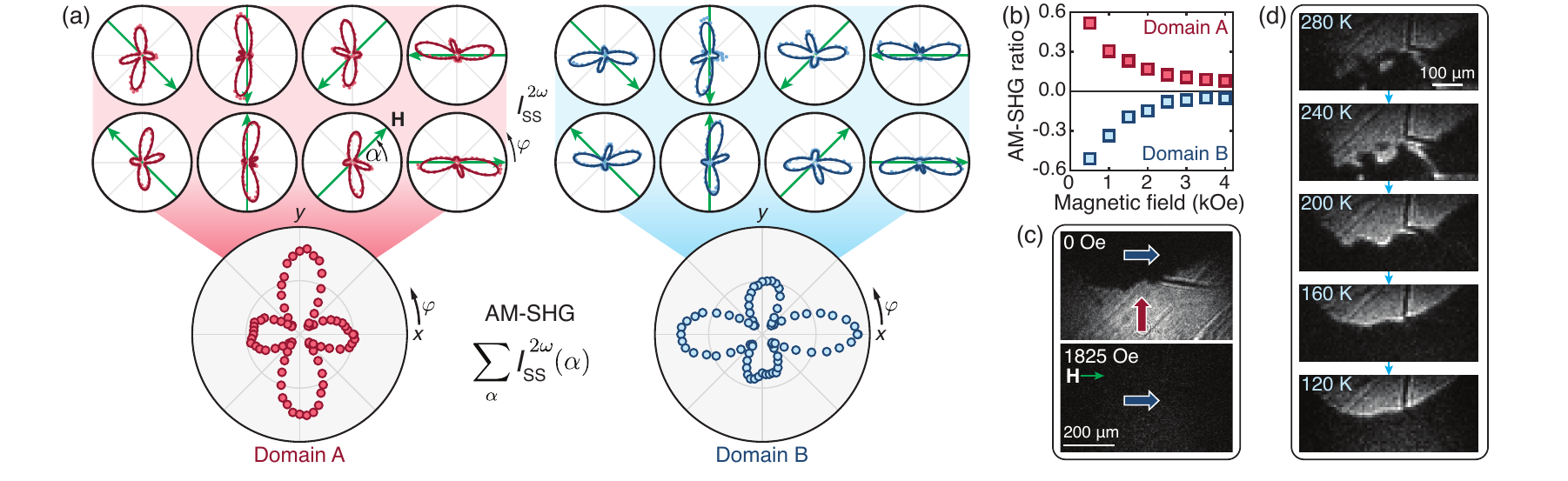}
\caption{\label{fig:m2} Evidence for uniaxial in-plane anisotropy. (a) \SiSo{} RA-SHG patterns at \SI{295}{\kelvin} on domains A and B with $|H| = 1$ kOe for different $\alpha$ (from \ang{0} to \ang{315} in \ang{45} steps). AM-SHG pattern (bottom) is produced by summing all eight RA patterns. (b) $|H|$-dependence of AM-SHG ratio on each domain, defined as $(I^\text{sum}_\text{y}-I^\text{sum}_\text{x})/I^\text{sum}_\text{min}$, where $I^\text{sum}_\text{x} (I^\text{sum}_\text{y})$ are AM-SHG lobe intensities along $x$ ($y$) and $I^\text{sum}_\text{min}$ is the smaller of $I^\text{sum}_\text{x}$ or $I^\text{sum}_\text{y}$. Higher absolute ratio values correspond to larger deviation from $C_4$ symmetry. (c) SHG images at $H = 0$ Oe (top) and $H = 1825$ Oe (bottom) at \SI{295}{\kelvin} under horizontal excitation polarization. $H$-field and magnetization directions are indicated by arrows. (d) SHG images showing \ang{90} domain wall temperature dependence. Dark lines correspond to terrace steps. Acquired with vertical polarization.}
\end{figure*}

A typical white light image of cleaved \ce{Sr2Cu3O4Cl2} shows a smooth surface except for a few lines from cleavage terraces (Fig.~\ref{fig:m1}f). Contrast between \ang{90} AFM domains is achieved using wide-field polarized SHG imaging at normal incidence. Under horizontal excitation polarization \cite{SI}, regions with $\mathbf{M}$ along the $\pm y$ ($\pm x$) direction appear bright (dark). An SHG image captured over the same field of view at $T=\SI{295}{\kelvin}$ shows clear bright and dark regions spanning hundreds of microns (Fig.~\ref{fig:m1}g). By collecting oblique incidence RA patterns at different locations throughout the imaged area (Fig.~\ref{fig:m1}h), we find that the entire bright (dark) region corresponds to a single AFM domain with $\mathbf{M}$ oriented along $-y$ ($-x$). The realization of only two out of four possible domain orientations is observed across multiple crystals. By repeating these measurements following multiple thermal cycles across \tnone{} and under different orientations of the crystal in the laboratory frame, we report two main phenomena \cite{SI}. First, the location of \ang{90} domain walls is largely reproducible, suggesting pinning to structural features. Second, the direction of $\mathbf{M}$ within the bright (dark) domain is fixed along either the $+y$ ($+x$) direction or the $-y$ ($-x$) direction, depending on the orientation of the crystal relative to the Earth's magnetic field. Anti-phase domains with \ang{180} walls are removed even by the weak field of Earth upon cooling below \tnone. These observations suggest that a particular AFM configuration is selected through an interplay of the Earth's field with an underlying uniaxial magnetic anisotropy along the $y$ ($x$) axis in the bright (dark) domain.  

The presence of uniaxial anisotropy can be probed using anisotropic magneto-SHG (AM-SHG), where RA patterns are measured under different applied in-plane magnetic field ($\mathbf{H}$) directions ($\alpha$) \cite{Seyler2020-wc}. Figure~\ref{fig:m2}a shows \SiSo{} RA patterns from two \ang{90} domains for different $\alpha$ with $H=1$ kOe. The AM-SHG patterns, obtained by summing RA patterns over $\alpha$, exhibit a clear two-fold rotational symmetry ($C_2$) characteristic of uniaxial anisotropy (Fig.~\ref{fig:m2}a), with the axis differing by \ang{90} for the two domains. These data confirm the presence of a domain-dependent in-plane uniaxial magnetic anisotropy. A low field was necessary for this measurement because for $H >$ 3 kOe, a spin rotation transition occurs for $\mathbf{H}$ along $\langle 100 \rangle$ \cite{Chou1997-fs,Kastner1999-ka}, which obscures the $C_2$ contribution to the AM-SHG patterns (Fig.~\ref{fig:m2}b). Field-dependent SHG imaging shows that a domain can be reoriented by \ang{90} at sufficiently high $H$ (Fig.~\ref{fig:m2}c), which appears to occur through the growth and merger of smaller domains \cite{SI}. Combined with the thermal cycling results, the data suggest that the structural symmetry is lower than tetragonal above \tnone. Structural domains may arise from previously unresolved high-temperature orthorhombic distortions, which impose spatially nonuniform uniaxial anisotropy below \tnone. The system may also be subject to extrinsic stresses from crystallographic defects as well as intrinsic stresses that are expected to accompany AFM order in finite crystals \cite{SI,Gomonay2011-im}.

The AFM domain distribution is dictated primarily by competition between the uniaxial anisotropy and \cuone{}-\cuone{} spin exchange energies. Although anisotropy is much weaker than exchange \cite{SI}, a \ang{90} AFM domain wall can nevertheless form along structural domain boundaries because the exchange energy cost scales with the wall area, whereas the anisotropy energy saved scales with the domain volume. Just below \tnone, we observe fragmentary AFM domains with rough edges (Fig.~\ref{fig:m2}d), likely conforming to an underlying distribution of structural domains. Upon further cooling, fragments merge to form larger AFM domains with smoother walls, indicating an increasing exchange contribution relative to the anisotropy that is possibly driven by changes in the ordered moment magnitude and temperature-dependent anisotropy. From \SI{160}{\kelvin} down to \SI{100}{\kelvin}, the domain boundaries remain largely stable. This general trend is consistent across multiple samples \cite{SI}.

\begin{figure*}[!htb]
\includegraphics{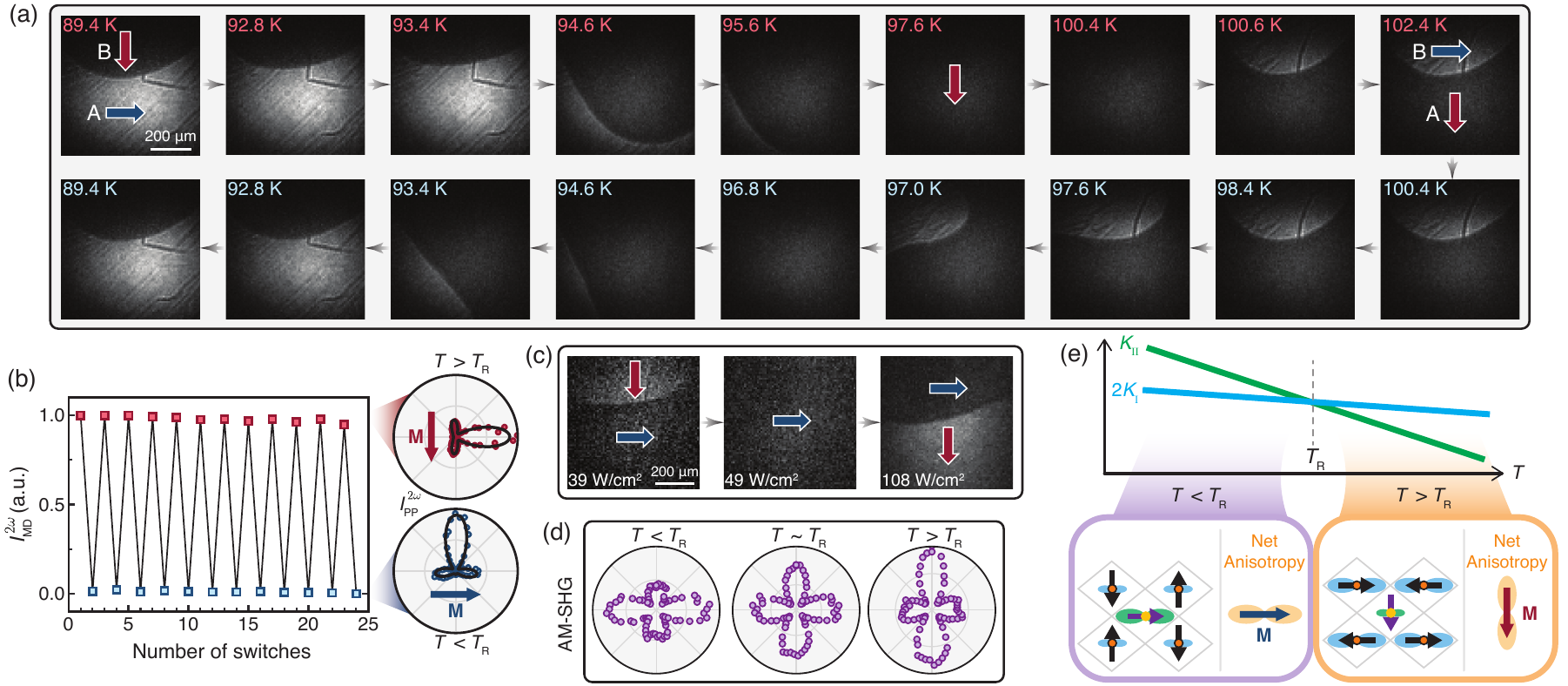}
\caption{\label{fig:m3} Thermally driven domain reorientation transition. (a) SHG images acquired at select temperatures upon warming (top) and cooling (bottom). Arrows indicate $\mathbf{M}$. Excitation was vertically polarized. Similar behavior was confirmed in a second sample \cite{SI}. (b) MD SHG intensity on domain A under repeated thermal switching through heating to \SI{120}{\kelvin} (red) and cooling to \SI{80}{\kelvin} (blue). Corresponding \PP{} patterns shown on right. (c) SHG images acquired at three different excitation intensities for \SI{95.5}{\kelvin}. (d) AM-SHG patterns acquired at $\treor-\SI{10}{\kelvin}$, \treor, and $\treor+\SI{5}{\kelvin}$ with $H = 1180$ Oe from a single domain. (e) Schematic of competing magnetic anisotropies model. The graph depicts how \cuone{} and \cutwo{} in-plane spin anisotropies, labelled as $K\textsubscript{I}$ and $K\textsubscript{II}$ respectively, may vary around \treor. The factor of two in front of $K\textsubscript{I}$ arises because there are twice as many \cuone{} as \cutwo{} per unit cell. In the crystal schematics, blue (green) lobe sizes and orientations depict the direction and strength of \cuone{} (\cutwo) spin anisotropy above and below \treor{} for domain A. Domain B is similar but rotated by \ang{90}.}
\end{figure*}

Figure~\ref{fig:m3}a shows the evolution of a typical \ang{90} AFM domain upon further cooling below \SI{100}{\kelvin}. Remarkably, within \SI{1}{\kelvin} around $\treor = \SI{97}{\kelvin}$, the domain wall is rapidly expelled from the sample---realizing a global single-domain state---and then reappears and snaps back into its original position with swapped bright and dark regions. This behavior is completely reversible upon re-heating through $\treor$ with slight thermal hysteresis. Local RA measurements confirm that $\mathbf{M}$ reorients by \ang{90} within each domain across \treor, with the two domains effectively swapping $\mathbf{M}$. While a previous study observed magnetization anomalies and inferred changes in relative domain sizes in this temperature regime \cite{Parks2001-hr}, the domain reorientation at \treor{} has remained hidden until this work. This reorientation transition enables repeated deterministic \ang{90} switching of the local AFM order parameter simply by cycling the cryostat temperature through \treor{} (Fig.~\ref{fig:m3}b), or by fixing the cryostat temperature below \treor{} and changing the optical power (Fig.~\ref{fig:m3}c).

The AM-SHG pattern from a single AFM domain has $C_2$ symmetry just above \treor, becomes $C_4$ at \treor, and then recovers a $C_2$ form below \treor{} that is \ang{90} rotated from the high-temperature pattern (Fig.~\ref{fig:m3}d). This strongly suggests that the reorientation transition is driven by a change in sign of the uniaxial magnetic anisotropy at \treor. In Fig.~\ref{fig:m3}e, we propose a simple microscopic picture in which \cuone{} and \cutwo{} spins exhibit temperature-dependent in-plane uniaxial magnetic anisotropies ($K\textsubscript{I}$ and $K\textsubscript{II}$). An expression for the anisotropy energy of a single domain is ${E(T)=E_0+[2K\textsubscript{I}(T)-K\textsubscript{II}(T)]\sin^2\psi-K_4\cos(4\psi)}$, where $\psi$ is the angle between $\mathbf{M}$ and the $+x$ direction, and $K_4$ is a biaxial anisotropy term \cite{Yildirim1995-vr, Chou1997-fs, Kastner1999-ka}. The \cuone{} and \cutwo{} spins prefer a relative orientation of \ang{90}. If $K\textsubscript{I}$ and $K\textsubscript{II}$ have the same sign but different strength (depicted by blue and green lobes in Fig.~\ref{fig:m3}e), the two terms compete, with the larger of $2K\textsubscript{I}$ and $K\textsubscript{II}$ determining the sign of the net uniaxial anisotropy and resulting orientation of $\mathbf{M}$. We hypothesize that $K\textsubscript{I}$ and $K\textsubscript{II}$ exhibit different temperature dependencies and cross at \treor{}, driving the \ang{90} AFM reorientation. Since \cuone{} and \cutwo{} lie at nonequivalent lattice sites, it is reasonable that a distortion-dependent single-ion anisotropy \cite{Liu2014-gd} will differ for each ion in both its strength and temperature dependence. The uniaxial anisotropy may also microscopically involve two-ion terms such as anisotropic exchange and magnetic dipole-dipole coupling, which is beyond the scope of this work to disentangle. We further note that spin correlations within the \cutwo{} sublattice have been shown to onset near $T=\SI{100}{\kelvin}$ \cite{Kim2001-fz}, potentially inducing magnetoelastic deformations.

This phenomenon is reminiscent of the transition across the isotropic point of \ce{Fe3O4} \cite{Bickford1950-ft, Martin-Garcia2016-ym} and the Morin transition of \ce{\alpha-Fe2O3} \cite{Morin1950-vs}. In these cases, spin reorientation occurs when temperature-dependent anisotropy contributions, originating from different magnetic ions or anisotropy mechanisms \cite{Broese_van_Groenou1969-tt, Artman1965-ew}, compensate one another to drive an anisotropy term across zero \cite{Belov1976-lt}. Our observations in \ce{Sr2Cu3O4Cl2} are distinguished from other temperature-dependent spin reorientation transitions in that the domain distribution is preserved, with the underlying distortions holding a memory of the domain structure while the anisotropy sets the spin orientation. Moreover, in \ce{Sr2Cu3O4Cl2}, the transition has been difficult to discern using bulk-averaged probes because it involves domain-dependent spin reorientation as opposed to a global change in easy axis.

\begin{figure}[!htb]
\includegraphics{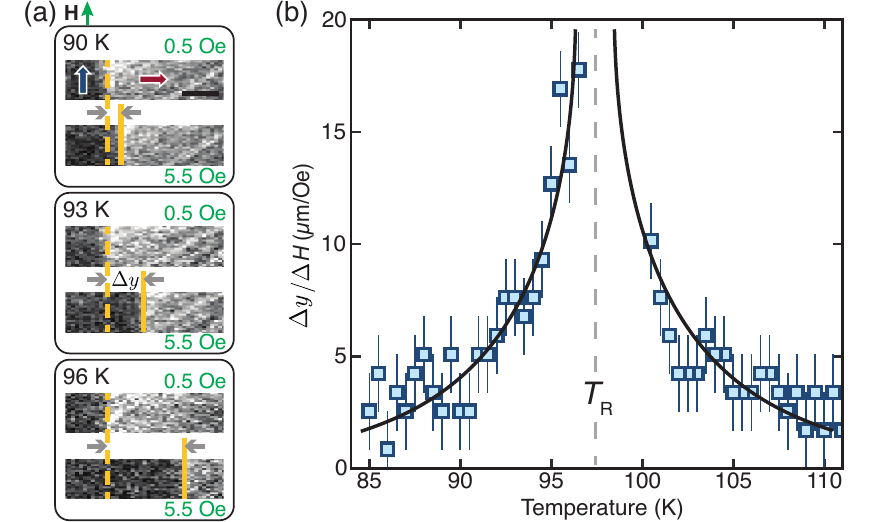}
\caption{\label{fig:m4} Divergence of the domain wall susceptibility. (a) SHG images displaying domain wall movement, $\Delta y$, when $H$ increases from \SI{0.5}{Oe} to \SI{5.5}{Oe} ($\Delta H = \SI{5}{Oe}$) at three different temperatures. Yellow dashed and solid vertical lines mark domain wall positions at \SI{0.5}{Oe} and \SI{5.5}{Oe} respectively. Scale bar, \SI{40}{\um}. Vertical pixel sizes are compressed twofold and images are rotated \ang{90} clockwise relative to other figures. (b) Temperature dependence of domain wall motion per change in $H$-field around \treor. Black lines are guides to the eye. Error bars are determined by uncertainty in domain wall horizontal positions.}
\end{figure}

Near \treor, the AFM domain walls become exceptionally responsive to small $H$. Figure~\ref{fig:m4}a illustrates the change in position ($\Delta y$) of the \ang{90} domain wall when $H$ is varied from 0.5 Oe to 5.5 Oe along the $x$ direction. As the temperature varies from \SI{90}{\kelvin} to \SI{96}{\kelvin}, $\Delta H$ has an increasingly large effect on domain wall motion. Since wall motion along $y$ is nearly uniform, the change in magnetization along $x$ is proportional to $\Delta y$, hence $\Delta y/\Delta H$ measures domain wall susceptibility. By repeating this experiment at many temperatures, we identify a striking divergence in the domain wall susceptibility at \treor, consistent with conclusions drawn from low-field magnetometry \cite{Parks2001-hr, SI}. As the net uniaxial anisotropy crosses zero, \ang{90} domain walls become energetically unfavorable and are easily expelled by small $H$. High domain wall tunability near \treor{} may be leveraged to prepare large AFM domains of a desired orientation \cite{SI,Gomonay2021-dq}.

Our approach to locally readout AFM states, globally image AFM domain walls, and deterministically switch \ang{90} domains in a cuprate Mott insulator augments existing AFM detection and manipulation schemes in other material classes, such as magnetoelectric oxides, rare-earth orthoferrites, and metallic alloys \cite{Song2018-wr}. Temperature-tunable anisotropy may be valuable for domain wall engineering, spin-superfluidity experiments \cite{Sonin2010-kz}, and studies of intrinsic domain wall mobility \cite{Thomas2007-nz,Kim2017-hg,Caretta2018-jl}. Because \ce{Sr2Cu3O4Cl2} is sensitive to small changes in the microscopic parameters, especially around \treor, it may be amenable to AFM manipulation with various other techniques, including strain tuning and nonthermal optical control \cite{De_la_Torre2021-ce}.

\begin{acknowledgments}
We acknowledge helpful conversations with Daniel Silevitch and Patrick Lee. The SHG measurements were supported by an ARO PECASE award W911NF-17-1-0204. D.H. also acknowledges support for instrumentation from the David and Lucile Packard Foundation and from the Institute for Quantum Information and Matter (IQIM), an NSF Physics Frontiers Center (PHY-1733907). K.L.S. acknowledges a Caltech Prize Postdoctoral Fellowship. A.R thanks the Zuckerman Foundation, and the Israel Science Foundation (grant No. 1017/20). The work at Stanford and SLAC (crystal growth and sample characterization) was supported by the U.S. Department of Energy (DOE), Office of Science, Basic Energy Sciences, Materials Sciences and Engineering Division, under contract DE-AC02-76SF00515.
\end{acknowledgments}


%

\newpage
\pagenumbering{gobble}
\begin{figure}
   \vspace*{-2cm}
   \hspace*{-2cm}
    \centering
    \includegraphics[page=1]{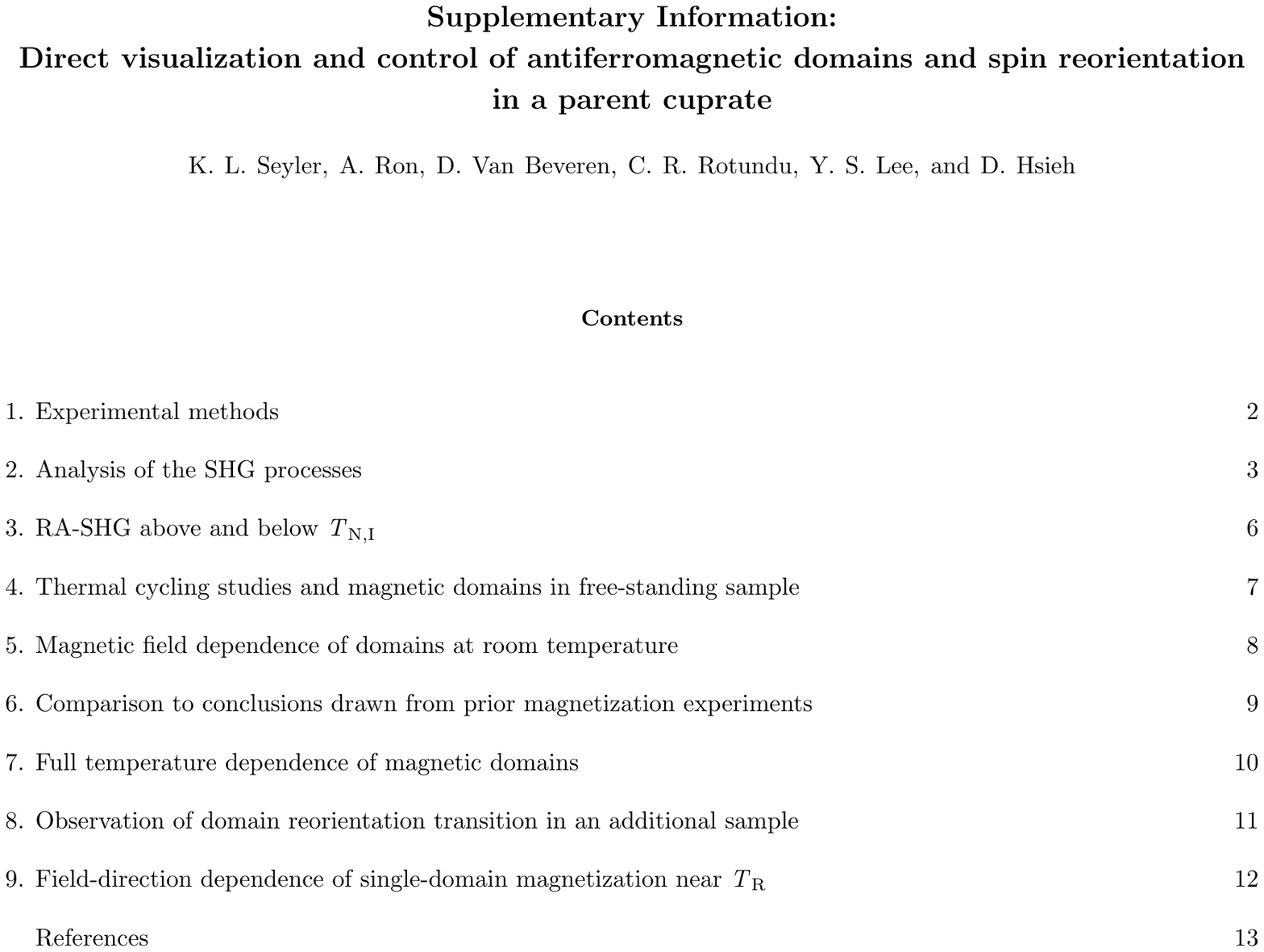}
    \caption{Caption}
    \label{fig:my_label}
\end{figure}
\begin{figure}
   \vspace*{-2cm}
   \hspace*{-2cm}
    \centering
    \includegraphics[page=2]{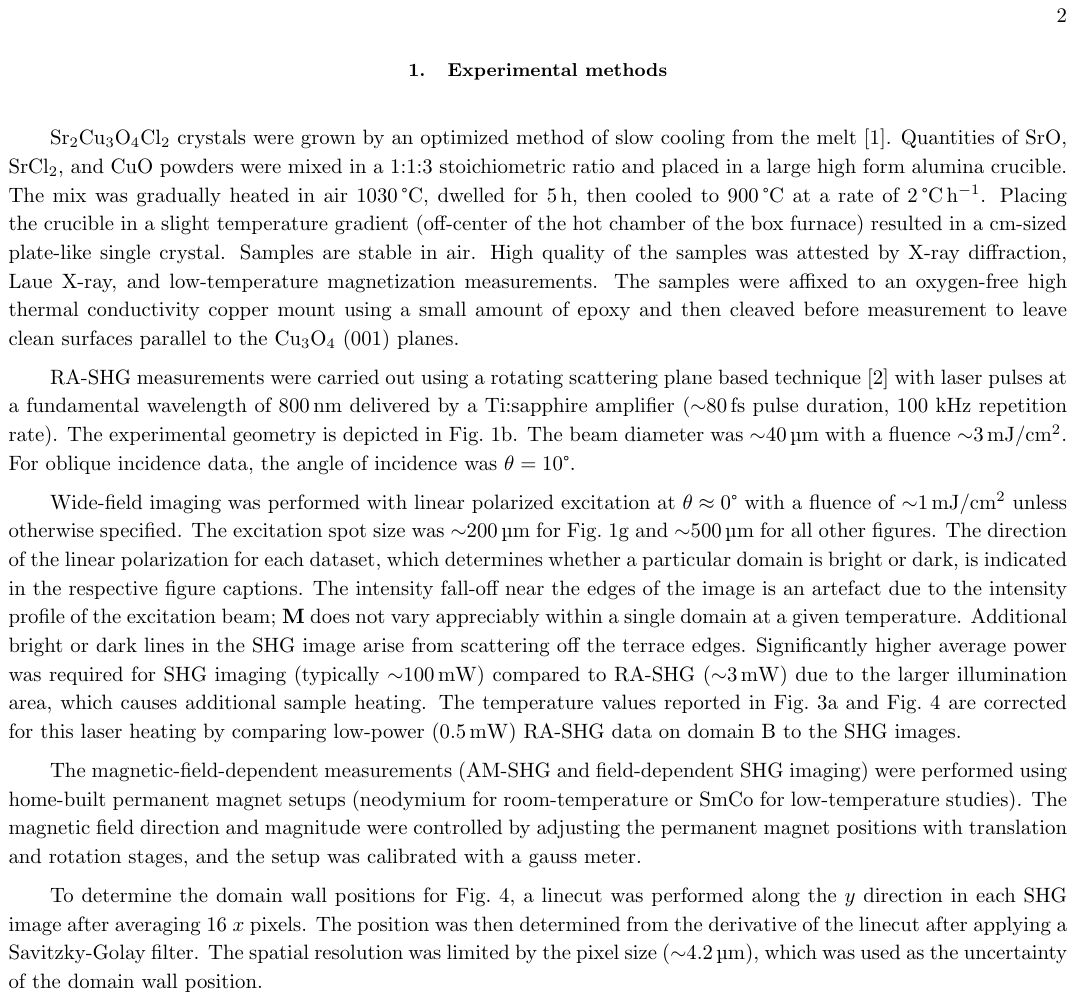}
    \caption{Caption}
    \label{fig:my_label}
\end{figure}
\begin{figure}
   \vspace*{-2cm}
   \hspace*{-2cm}
    \centering
    \includegraphics[page=3]{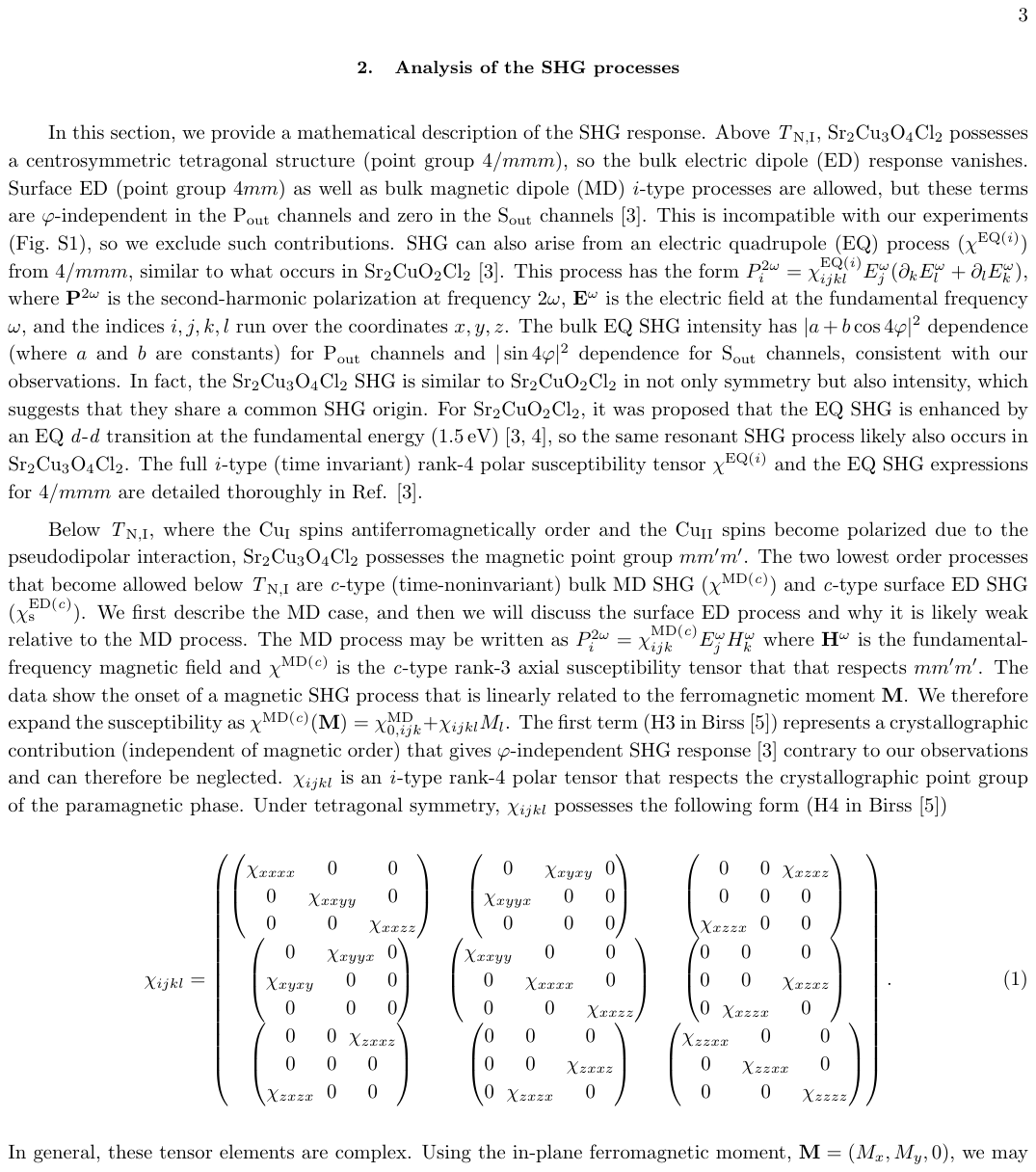}
    \caption{Caption}
    \label{fig:my_label}
\end{figure}
\begin{figure}
   \vspace*{-2cm}
   \hspace*{-2cm}
    \centering
    \includegraphics[page=4]{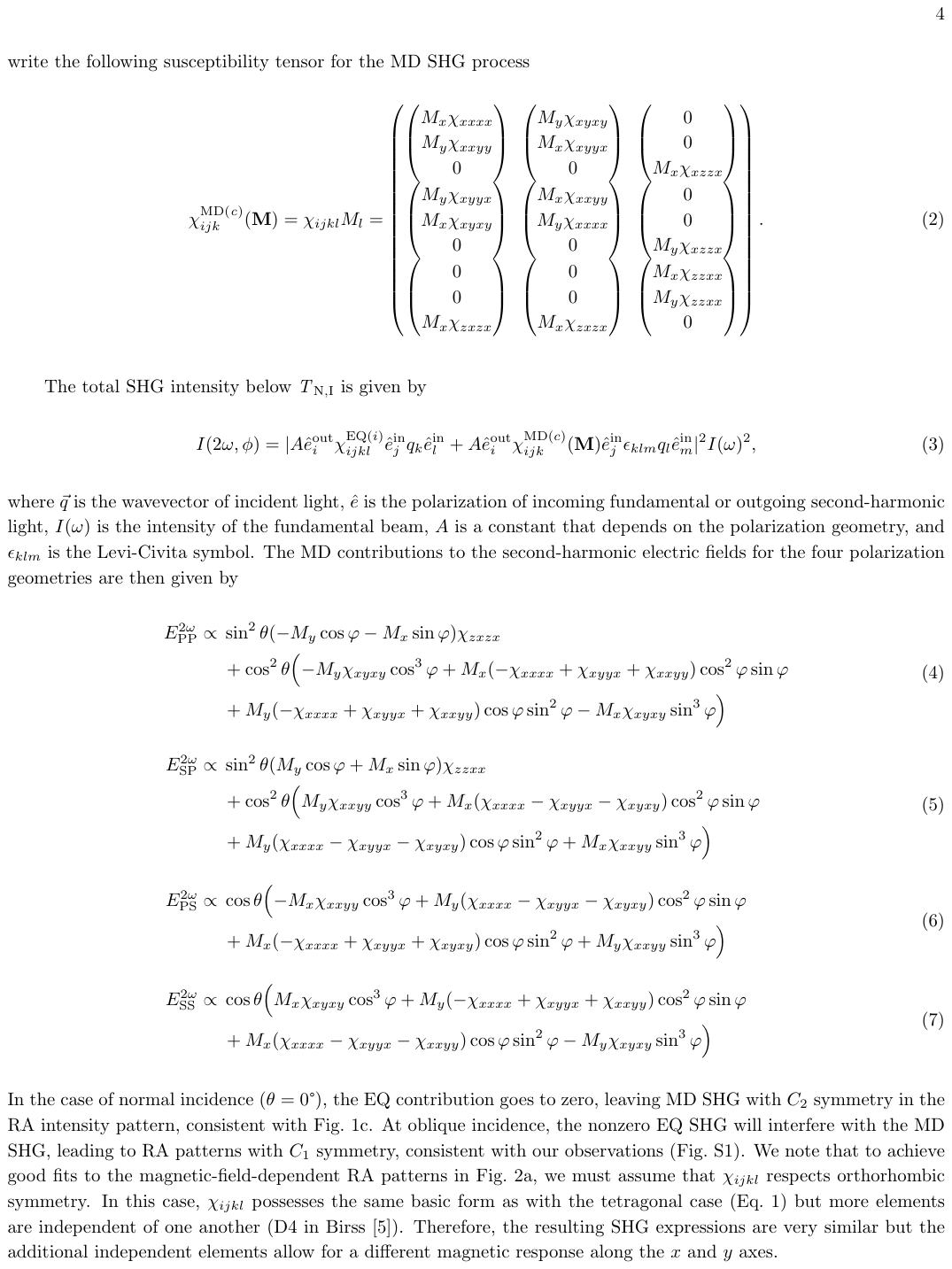}
    \caption{Caption}
    \label{fig:my_label}
\end{figure}
\begin{figure}
   \vspace*{-2cm}
   \hspace*{-2cm}
    \centering
    \includegraphics[page=5]{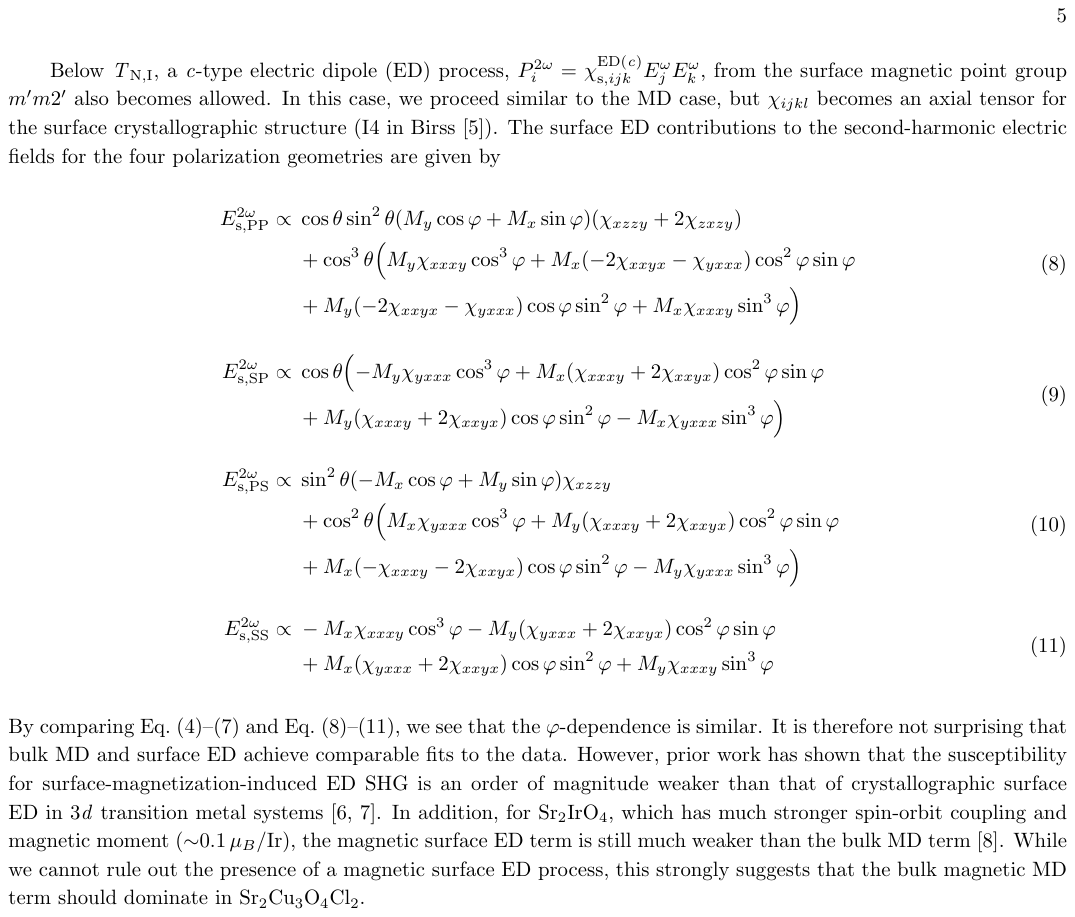}
    \caption{Caption}
    \label{fig:my_label}
\end{figure}
\begin{figure}
   \vspace*{-2cm}
   \hspace*{-2cm}
    \centering
    \includegraphics[page=6]{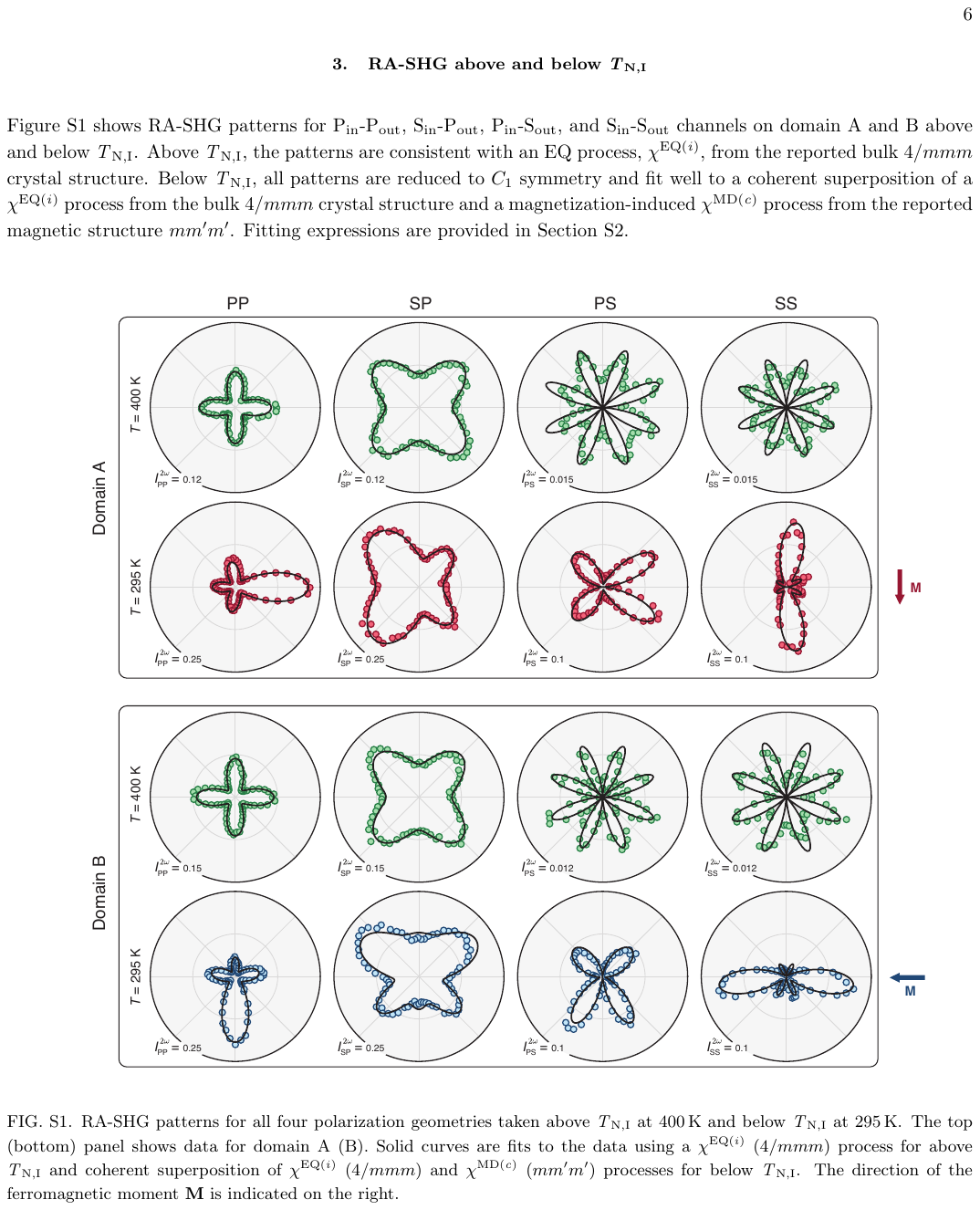}
    \caption{Caption}
    \label{fig:my_label}
\end{figure}
\begin{figure}
   \vspace*{-2cm}
   \hspace*{-2cm}
    \centering
    \includegraphics[page=7]{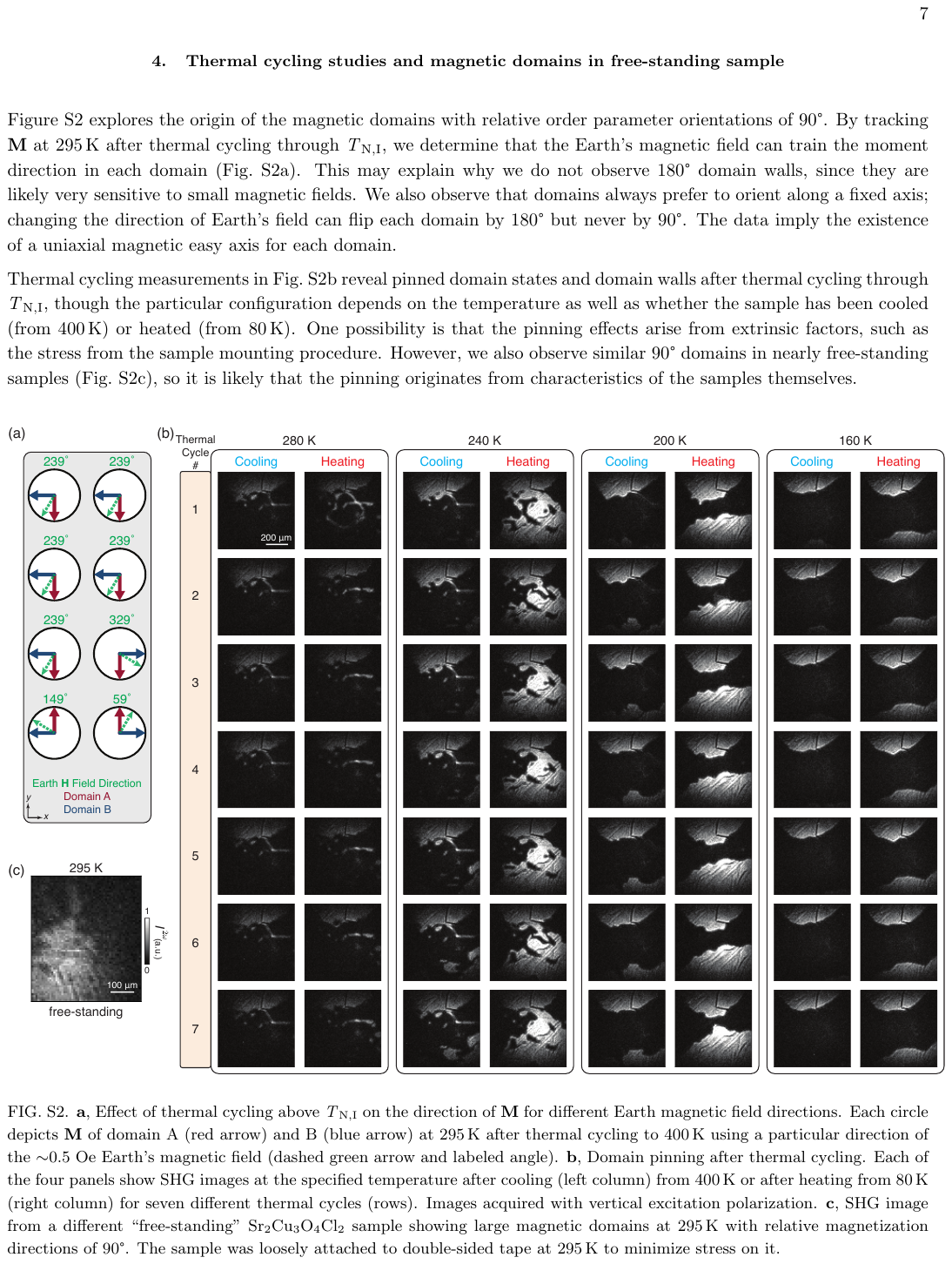}
    \caption{Caption}
    \label{fig:my_label}
\end{figure}
\begin{figure}
   \vspace*{-2cm}
   \hspace*{-2cm}
    \centering
    \includegraphics[page=8]{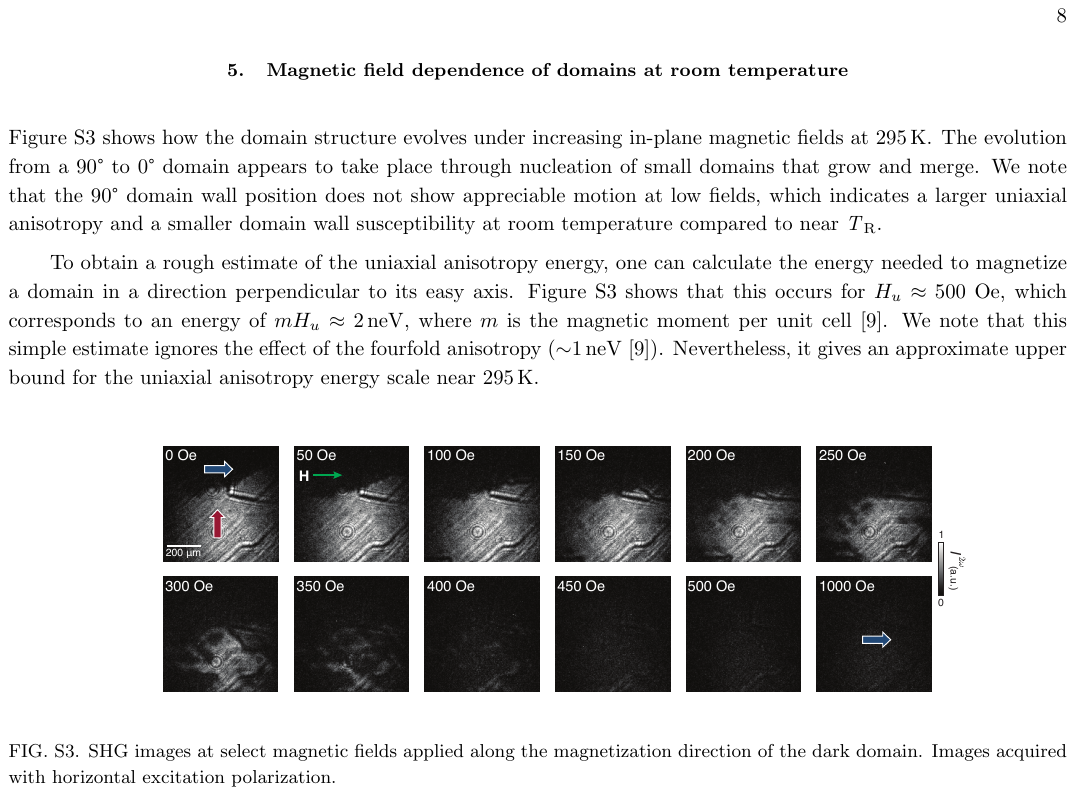}
    \caption{Caption}
    \label{fig:my_label}
\end{figure}\begin{figure}
   \vspace*{-2cm}
   \hspace*{-2cm}
    \centering
    \includegraphics[page=9]{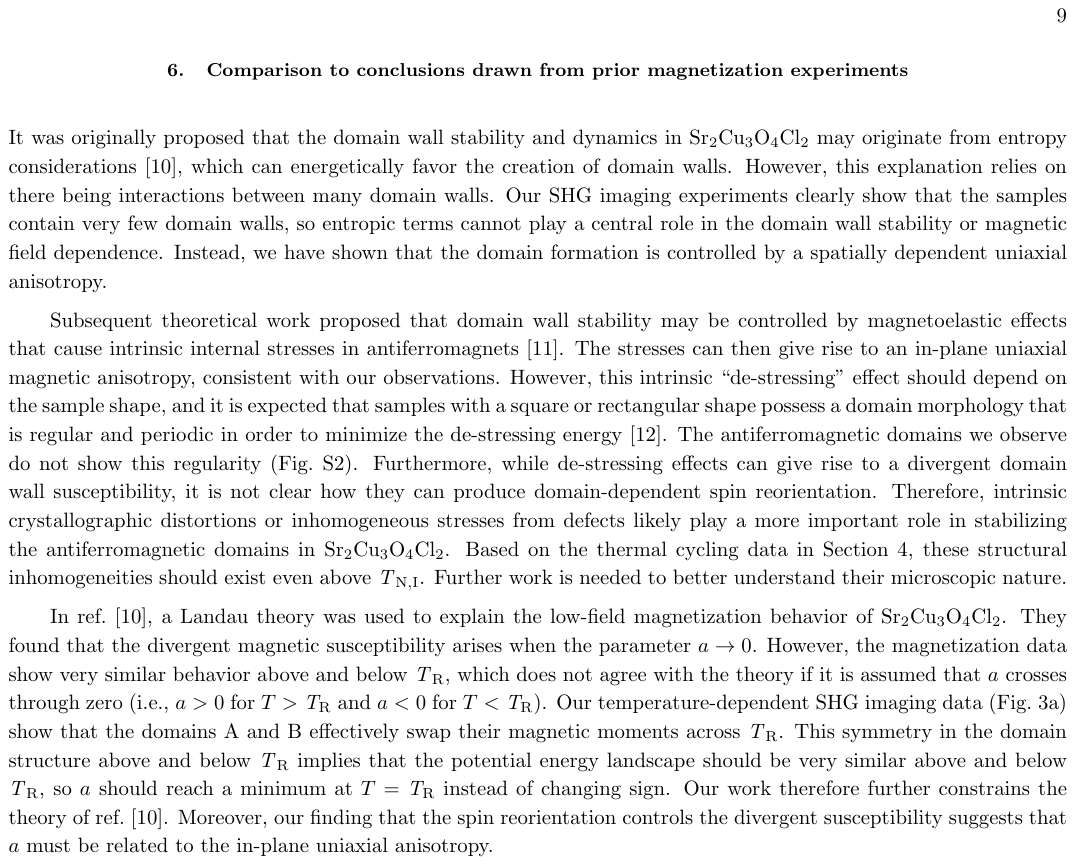}
    \caption{Caption}
    \label{fig:my_label}
\end{figure}
\begin{figure}
   \vspace*{-2cm}
   \hspace*{-2cm}
    \centering
    \includegraphics[page=10]{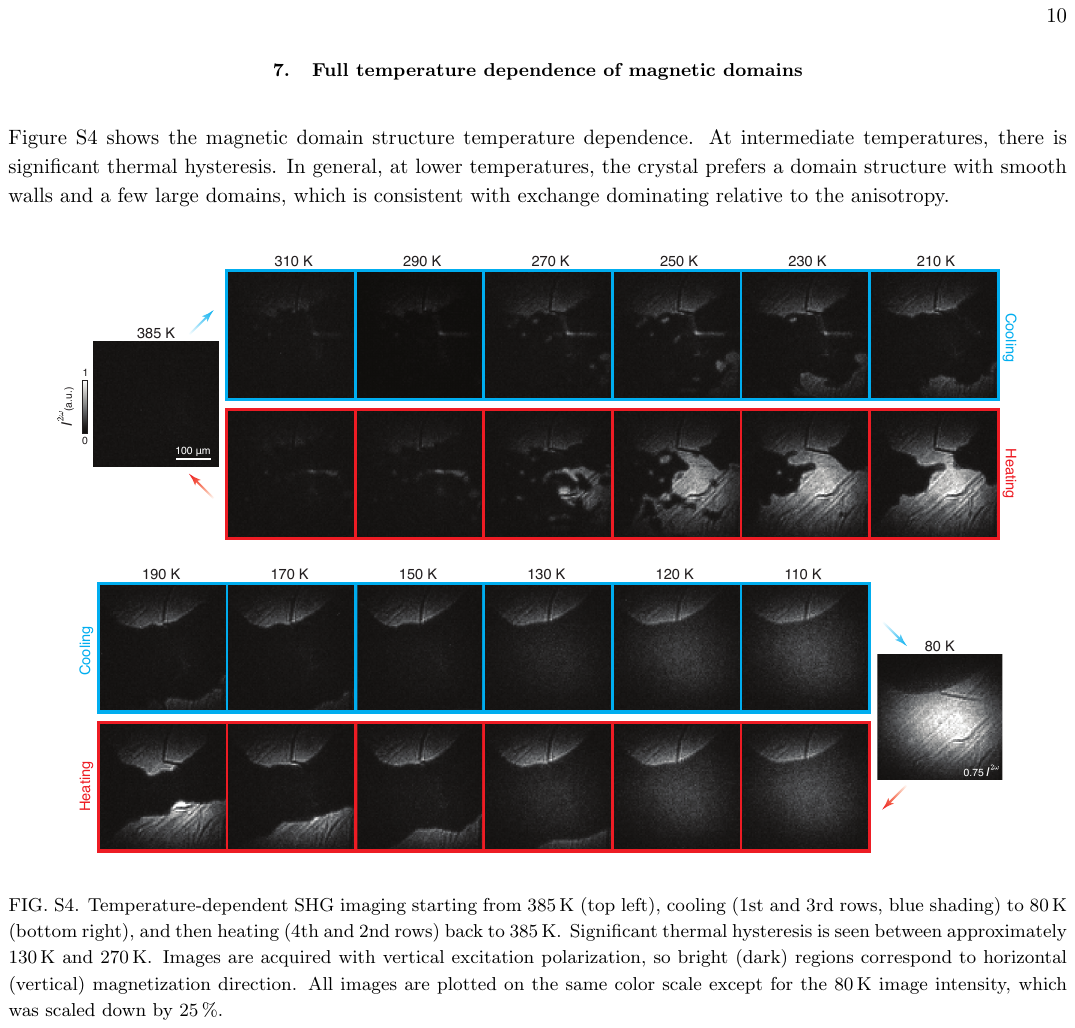}
    \caption{Caption}
    \label{fig:my_label}
\end{figure}
\begin{figure}
   \vspace*{-2cm}
   \hspace*{-2cm}
    \centering
    \includegraphics[page=11]{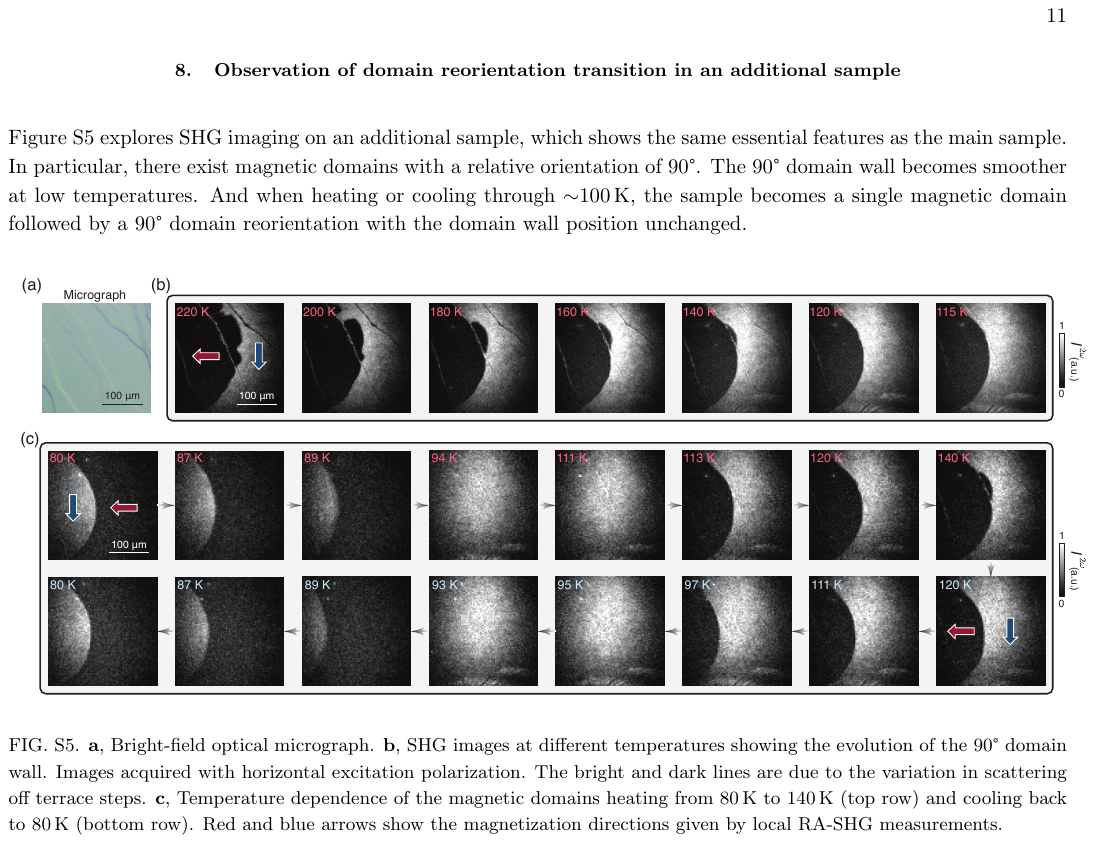}
    \caption{Caption}
    \label{fig:my_label}
\end{figure}
\begin{figure}
   \vspace*{-2cm}
   \hspace*{-2cm}
    \centering
    \includegraphics[page=12]{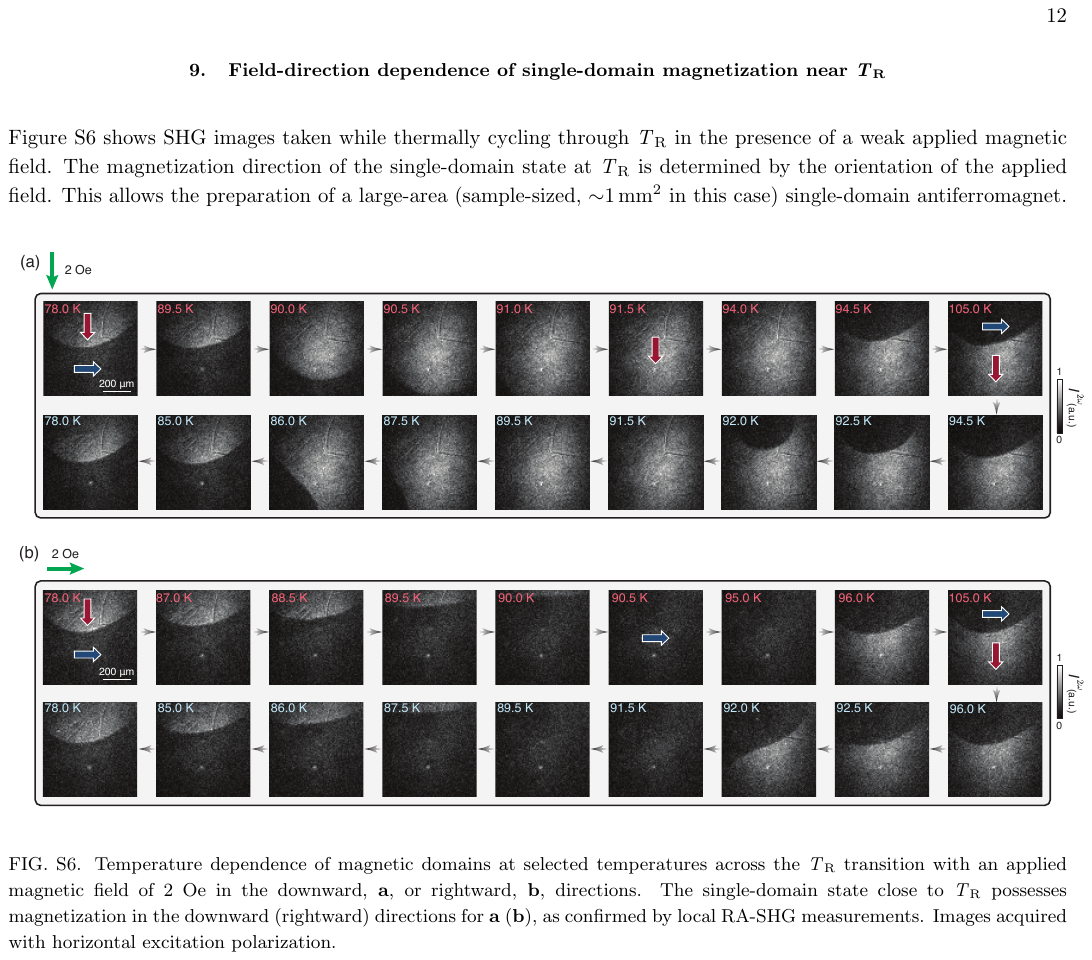}
    \caption{Caption}
    \label{fig:my_label}
\end{figure}
\begin{figure}
   \vspace*{-2cm}
   \hspace*{-2cm}
    \centering
    \includegraphics[page=13]{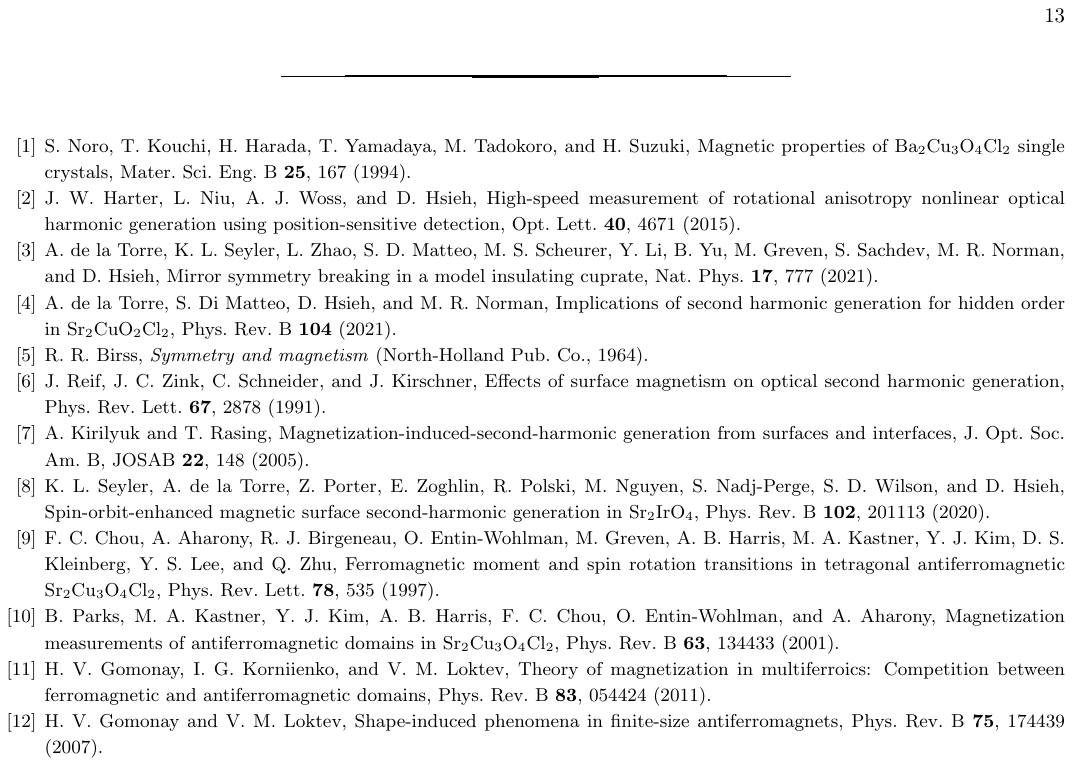}
    \caption{Caption}
    \label{fig:my_label}
\end{figure}

\end{document}